\documentclass{llncs}
\usepackage{lmodern}
\usepackage[utf8]{inputenc}
\usepackage{color}
\usepackage{soul}
\usepackage{graphicx}
\usepackage{xspace}
\usepackage{multirow}
\usepackage{url}
\usepackage{amsmath,amssymb,amsfonts}
\usepackage{microtype,mathtools,mathrsfs}
\usepackage{hyperref}
\hypersetup{pdfstartview=XYZ}

\newtheorem{observation}{Observation}

\renewcommand{\leq}{\leqslant}

\renewcommand{\geq}{\geqslant}

\newcommand{\Turn}{\texttt{Turn}}
\newcommand{\DR}{\texttt{DR}}

\newcommand{\pebble}{\texttt{Pebble}}

\usepackage{graphicx}
\graphicspath{ {images/} }

\begin{document}

\frontmatter          
%
%

\title{Fast Rendezvous on a Cycle by Agents \\ with Different Speeds}
\author{Ofer Feinerman\inst{1}
\and
Amos Korman\inst{2}
\and
Shay Kutten\inst{3}
\and
Yoav Rodeh\inst{4}
}
\institute{The Shlomo and Michla Tomarin Career Development Chair,  Weizmann Institute of Science, Rehovot, Israel.
 \and
CNRS and University Paris Diderot, Paris, France.
\and Faculty of IR\&M, Technion, Haifa 32000, Israel.
\and
Jerusalem College of Engineering.
}
\date{}
\maketitle

\begin{abstract}
The difference between the speed of the actions  of different processes is typically considered as an obstacle that makes the achievement of cooperative goals more difficult. In this work, we aim to highlight  potential  {\em benefits} of such asynchrony phenomena to tasks involving symmetry breaking.
Specifically, in this paper, identical (except for their speeds) mobile agents are placed at arbitrary locations on a
(continuous) cycle of length $n$ and use their speed difference in order to rendezvous fast.
We normalize the speed of the slower agent to be 1, and fix the speed of the faster agent to be some  $c>1$. (An agent does not know whether it is the slower agent or the faster one.)
The  straightforward {\em distributed-race ($\DR$)} algorithm is the one in which both agents simply start walking until rendezvous is achieved. It is easy to show that, in the worst case, the rendezvous time of $\DR$ is $n/(c-1)$. Note that in the interesting case, where $c$ is very close to 1 (e.g., $c=1+1/n^k$), this bound  becomes huge. Our first result is a lower bound showing that, up to a multiplicative  factor of 2, this bound is unavoidable, even in a model that allows agents to leave arbitrary marks (the {\em white board} model), even assuming  sense of direction, and even assuming $n$ and $c$ are known to agents.
That is, we show that under such assumptions,  the rendezvous time of any algorithm is at least  $\frac{n}{2(c-1)}$ if  $c\leq 3$ and slightly larger (specifically, $\frac{n}{c+1}$) if $c>3$.
We then manage to construct an algorithm that precisely matches the lower bound for the case $c\leq 2$, and almost matches it when $c>2$. Moreover, our algorithm performs under weaker assumptions than those stated above, as it does not assume sense of direction, and
it allows agents to leave only a single mark (a pebble) and only at the place where they start the execution. Finally, we investigate the setting in which no marks can be used at all, and show tight bounds for $c\leq 2$, and almost tight bounds for $c>2$.
\end{abstract}

\clearpage
\section{Introduction}

\subsection{Background and Motivation}

The difference between the speed of the actions  of different entities is typically considered disruptive in real computing systems.
In this paper, we  illustrate some advantages of such phenomena in  cases where the difference remains {\em fixed} throughout the execution\footnote{Advantages can also be exploited in cases where the difference in speed follows some stochastic distribution, however, in this initial study, we focus on the simpler fully deterministic case. That is, we assume a speed heterogeneity that is arbitrary but fixed throughout the execution.}.
  We demonstrate the  {\em usefulness} of this manifestation of asynchrony to tasks involving symmetry breaking. More specifically, we show how two mobile agents, identical in every aspect save their speed, can lever their speed difference in order to achieve fast rendezvous.

Symmetry breaking  is a major issue in distributed computing that is completely absent from traditional sequential computing. Symmetry can often prevent different processes from reaching a common goal. Well known examples include leader election~\cite{angluin}, mutual exclusion  \cite{dijkstra}, agreement
\cite{anon-agree,anon-robot} and renaming \cite{renaming}. To address this issue, various differences between processes are exploited. For example, solutions for leader election often rely on unique identifiers assumed to be associated with each entity (e.g., a process) \cite{angluin}. Another example of a difference is the location of the entities in a network graph. Entities located in different parts of a non-symmetric graph can use this knowledge in order to behave differently; in such a case, a leader can be elected even without using unique identifiers \cite{kameda}.
If no differences exist, breaking symmetry deterministically becomes impossible (see, e.g., \cite{angluin,yu-yung}) and one must resort to randomized algorithms, assuming~that different entities can draw different random bits \cite{itai-rodeh}.

We consider mobile agents aiming to rendezvous. See, e.g.,~\cite{BCGIL10,CILP11,kranakis-book,kranakis-survey,santoro-ring,yu-yung}.
As is the case with other symmetry breaking problems,
it is well known that if the agents are completely identical then rendezvous is, in some cases, impossible.
In fact, a large portion of the research about rendezvous dealt with identifying the conditions under which rendezvous was possible, as a result of some asymmetries.
Here, the fact that agents have different speeds implies that the mere feasibility of rendezvous is trivial, and our main concern is therefore the time complexity, that is, the time to reach a rendezvous.
More specifically,
we study the case where the agents are identical except for the fact that they have different speeds of motion.
Moreover, to isolate the issue of the speed difference, we remove other possible differences between the agents. That is, the agents are assumed to be anonymous. To  avoid solutions of the kind of \cite{kameda}, that are based on the underlying graph being asymmetric, we consider
a symmetric topological object, that is, specifically, a cycle topology. We denote by $n$ the length of the cycle.

\subsection{The Model and the Problem}

\paragraph{The problem of rendezvous on a cycle:} Consider two identical deterministic {\em agents} placed on
a cycle of length $n$ (in some distance units). To ease the description, we name these agents $A$ and $B$ but these names
 are not known to the agents.  Each agent is initially placed in some location
on the cycle by an adversary and both agents  start the execution of the algorithm simultaneously. An agent can move on the cycle at any direction. Specifically, at any given point in time, an agent can decide to either start moving, continue in the same direction, stop, or change its direction.
 The agents' goal is to {\em rendezvous}, namely, to get to be co-located  somewhere on the cycle\footnote{In a sense, this rendezvous problem is also similar to the {\em cow-path} problem, see, e.g.,~\cite{BCR91}. Here, the agents (the cow and the treasure she seeks to find) are  both mobile (in the cow-path problem only one agent, namely, the cow, is mobile). It was shown in \cite{BCR91} that if the cow is initially located at distance $D$ from the treasure on the infinite line then the  time to find the treasure can be $9D$, and that 9 is the best multiplicative constant (up to lower order terms in $D$).}. We consider
 continuous movement, so this rendezvous can occur at any location along the cycle. An agent can detect the presence of another agent at its location and hence detect  a rendezvous. When agents detect a rendezvous, the rendezvous task is considered completed.

\paragraph{Orientation issues:}
We distinguish between two models based on orientation. The first assumes that  agents have the {\em sense of direction}~\cite{santoro-sense-of-direction}, that is, we assume that the agents can distinguish  clockwise from the anti-clockwise. In the second model, we do not assume this orientation assumption. Instead, each agent has its own perception of which direction is clockwise and which is anti-clockwise, but there is no guarantee that these perceptions are globally consistent. (Hence, e.g., in this model, if both agents start walking in their own clockwise direction, they may happen to walk in opposite directions, i.e., towards each other).

 \paragraph{The {\em pebble}  and the {\em white board} models:}
  Although the agents do not hold any direct means of communication,  in some cases we do assume that an agent can leave marks in its current location on the cycle, to  be read later
  by itself and by the other agent. In the {\em pebble} model, an agent can
 mark its location by dropping a pebble~ \cite{bender-pebble,blum-kozen-pebble}.
Both dropping and detecting a pebble are local acts taking place only on the location occupied by the agent.
We note that in the case where pebbles can be dropped, our upper bound employs agents that drop a pebble only once and only at their initial location~\cite{presenter,baston-gal-mark-starting,santoro-ring}. On the other hand, our corresponding lower bound holds for any mechanism of (local) pebble dropping. Moreover, this lower bound holds also for the seemingly stronger '{\em white board}  model, in which an agent can change a memory associated with its current location such that it could later be read and further manipulated by itself or the other agent~\cite{KKM,white-board,white-board2}.

\paragraph{Speed:}
Each agent moves at the same fixed speed at all times;
the {\em speed} of an agent $A$, denoted $s(A)$, is the inverse of
 the time $t_{\alpha}$ it takes agent $A$ to traverse one unit of length.
 For agent $B$, the time $t_{\beta}$ and speed $s(B)$ are defined analogously.
Without loss of generality, we assume that agent $A$ is faster, i.e.,  $s(A)>s(B)$ but  emphasize that this is unknown to the agents themselves.
Furthermore, for simplicity of presentation, we normalize the speed of the slower agent $B$ to one, that is,
$s(B)=1$ and denote $s(A)= c$ where $c>1$.  We stress that the more interesting cases are when $c$ is a function of $n$ and arbitrarily close to 1 (e.g., $c=1+1/n^k$, for some constant $k$).
We assume that each agent has a pedometer that enables it to measure the distance it travels. 

In some cases, agents are assumed to posses some knowledge regarding $n$ and $c$; whenever used, this assumption will be mentioned explicitly.

\paragraph{Time complexity:}
The {\em rendezvous time} of an algorithm is defined as the worst case time bound until rendezvous, taken over all pairs of initial placements of the two agents on the cycle. Note, a lower bound for the rendezvous time that is established assuming  sense of direction holds trivially for the case where no sense of direction is assumed. All our lower bounds hold assuming sense of direction.

\paragraph{The {\em Distributed Race} ($\DR$) algorithm:}
 Let us consider a trivial algorithm, called  {\em Distributed Race} ($\DR$), in which an agent starts moving in an arbitrary direction, and continues to walk in that  direction until reaching rendezvous.
 Note that this algorithm does not assume knowledge of $n$ and $c$, does not assume sense of direction and does not leave marks on the cycle.
 The worst case for this algorithm is that both agents happen to walk on the same direction. Without loss of generality, assume this direction is clockwise. Let $d$ denote the  the clockwise distance from the initial location of $A$ to that of $B$.
The rendezvous time $t$   thus satisfies $t \cdot s(A)= t \cdot s(B)+d$.
Hence, we obtain the following.

\begin{observation}\label{DR}
The rendezvous time of $\DR$ is $d/(c-1)< n/(c-1)$.
\end{observation}
Note that in the cases where $c$ is very close to 1, e.g., $c=1+1/n^k$, for some constant $k$, the bound on the rendezvous time of $\DR$  is very large.

\subsection{Our Results}
Our first result is a lower bound showing that, up to a multiplicative approximation factor of 2, the bound of $\DR$ mentioned in Observation \ref{DR} is unavoidable, even in the white board model, even assuming  sense of direction, and even assuming $n$ and $c$ are known to agents.
That is, we show that under such assumptions,  the rendezvous time of any algorithm is at least  $\frac{n}{2(c-1)}$ if  $c\leq 3$ and slightly larger (specifically, $\frac{n}{c+1}$) if $c>3$.
We then manage to construct an algorithm that  matches the lower bound precisely for the case $c\leq 2$, and almost matches it when $c>2$.
Specifically, when $c\leq 2$, our algorithm runs in  time $\frac{n}{2(c-1)}$ and when $c>2$, the rendezvous time is~$n/c$ (yielding a $(2-\frac{2}{c})$-approximation when $2<c\leq 3$, and a $(\frac{c+1}{c})$-approximation when $c>3$). Moreover, our algorithm performs under weaker assumptions than those stated above, as it does not assume sense of direction, and allows agents to leave only a single mark (a pebble) and only at the place where they start the execution.

Finally, we investigate the setting in which no marks can be used at all, and show tight bounds for $c\leq 2$, and almost tight bounds for $c>2$.
Specifically, for this case, we establish a tight bound of $\frac{cn}{c^2 - 1}$ for the rendezvous time, in case agents have sense of direction. With the absence of sense of direction, the same lower bound of $\frac{cn}{c^2 - 1}$ holds, and
we obtain an algorithm matching this bound for the case $c\leq 2$, and  rather efficient algorithms for the case $c>2$. Specifically, the rendezvous time for  the case $c\leq 2$ is $\frac{cn}{c^2 - 1}$,  for the case $2<c\leq 3$, the rendezvous time is $\frac{2n}{c+1}$, and for the case $c>3$, the rendezvous time is  $\frac{n}{c-1}$.

\section{A Lower Bound for the White Board Model}
The following lower bound implies that $\DR$ is a 2-approximation algorithm, and it becomes close to optimal when $c$ goes to infinity.

\begin{figure}[ht]
\includegraphics[width=0.5\textwidth]{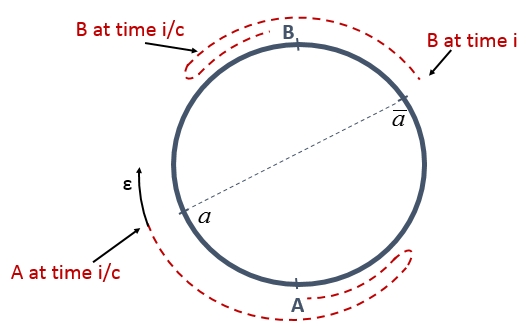}
\centering
\caption{For $B$ to be influenced by $A$'s last moves, it must be $\epsilon$-near to $A$ between times $i$ and 
$\frac{i}{c}$ but it must also be able to reach $\overline{a}$ at time $i$ by the induction hypothesis.}
\label{fig:lower}
\end{figure}

\begin{theorem}\label{lower}
Any rendezvous algorithm in the white board model requires at least $\max\{\frac{n}{2(c-1)}, \frac{n}{c+1} \}$ time, even assuming  sense of direction and even assuming $n$ and $c$ are known to the agents.
\end{theorem}
\begin{proof}
We assume that agents have  sense of direction; hence, both agents start walking at the same direction.
We first show that any algorithm in the white board model requires $\frac{n}{2(c-1)}$ time.
Consider the case that the adversary locates the agents at {\em symmetric locations} of the cycle,
i.e., they are at distance $n/2$ apart.
Now consider any algorithm used by the agents. Let us fix any $c'$ such that $1<c'<c$, and define the (continuous)  interval  $$I:=\left[0,\frac{nc'}{2(c-1)}\right].$$ For every (real) $i\in I$, let us define the following (imaginary) scenario $S_i$. In scenario $S_i$, each agent executes the algorithm for $i$ units of distance (not necessarily in the same direction) and terminates\footnote{We can think of this scenario as if each agent executes another algorithm $B$, in which it simulates precisely $i$ units of distance of the original algorithm and then terminates.}.
We claim that for every $i\in I$, the situation at the end of scenario $S_i$ is completely symmetric: that is, the white board at symmetric locations contain the same information and the two agents are at symmetric locations. We prove this claim by induction. The basis of the induction, the case $i=0$, is immediate.
Let us assume that the  claim holds for scenario $S_i$, for (real) $i\in I$, and consider scenario $S_{i+\epsilon}$, for any
positive $\epsilon$ such that $\epsilon~<~\frac{n}{4}(1-\frac{c'}{c}).$ Our goal is to show that the claim holds for scenario $S_{i+\epsilon}$\footnote{Note that  for some $i\in I$ and some $\epsilon~<~\frac{n}{4}(1-\frac{c'}{c})$, we may have that $i+\epsilon\notin I$. Our proof
will show that the claim for $S_{i+\epsilon}$ holds also in such cases. However, since we wish to show that the claim holds for $S_j$, where $j\in I$, we are not really interested in those cases, and are concerned only with the cases where $i+\epsilon \in I$ and $i \in I$.}.

Consider scenario $S_{i+\epsilon}$. During the time interval $[0,\frac{i}{c})$,  both agents perform the same actions as they do in the corresponding time interval in scenario $S_i$.
Let $a$ denote the location of agent $A$ at time $i/c$.
Now, during the time period $[\frac{i}{c},\frac{i+\epsilon}{c}]$, agent $A$ performs some movement all of which is done at distance at most $\epsilon$ from $a$ (during this movement  it may write information at various locations it visits); then, at time $\frac{i+\epsilon}{c}$, agent $A$ terminates.

Let us focus on agent $B$ (in scenario $S_{i+\epsilon}$) during the time period $[\frac{i}{c},i]$.
We claim that during this time period, agent~$B$ is always at distance at least $\epsilon$ from $a$.
Indeed, as long as it is true that agent $B$ is at distance at least $\epsilon$ from $a$, it performs the same actions as it does
in scenario $S_{i}$ (because it is unaware of any action made by  agent $A$ in  scenario~$S_{i+\epsilon}$, during the time  period $[\frac{i}{c},\frac{i+\epsilon}{c}]$). Therefore, if at some time $t'\in [\frac{i}{c},i]$, agent $B$ is (in scenario $S_{i+\epsilon}$) at distance less than $\epsilon$ from $a$ then the same is true also for scenario  $S_{i}$.  However, in scenario  $S_{i}$, by the induction hypothesis, agent $B$ was at time $i$ at $\bar{a}$, the symmetric location of $a$, that is, at distance $n/2$ from $a$. Thus, to get from a distance less than $\epsilon$ from $a$ to $\bar{a}$, agent $B$ needs to travel a distance of $n/2-\epsilon$, which takes
$n/2-\epsilon$ time (see figure \ref{fig:lower} for the argument up to this point). This is impossible since
$$i-i/c~\leq~\frac{nc'}{2c}~<~\frac{n}{2}-\epsilon,$$
where the first inequality follows from the definition of $I$ and the second follows from the definition of $\epsilon$.
It follows that during the time period from $[\frac{i}{c},i)$, agent $B$ behaves (in scenario~$S_{i+\epsilon}$) the same as it does in the corresponding time period  in scenario~$S_{i}$. Therefore, according to the induction hypothesis, 
 in  scenario~$S_{i+\epsilon}$, when completing $i$ units of distance, agent $B$ is at distance $n/2$ from where agent $A$ is after completing $i$ units of distance (recall, at that point agent $A$ is at $a$), and the cycle configuration (including the white boards) is completely symmetric. Now, since $\epsilon<n/4$, during the time period $[{i},{i+\epsilon}]$, agent $B$ is still at a distance more than $\epsilon$ from $a$ and  remains unaware of any action made  by agent $A$, during the time  period $[\frac{i}{c},\frac{i+\epsilon}{c}]$. (Similarly, agent $A$, during the time  period $[\frac{i}{c},\frac{i+\epsilon}{c}]$, is unaware of any action made  by agent $B$ during this time  period.)
Hence, at each time $i'\in [{i},{i+\epsilon}]$, agent $B$ takes the same action as agent $A$ in the corresponding time $i'/c$. This establishes the induction proof.
To sum up, we have just shown that for any $i\in I$,  the cycle configuration at the end of scenario $S_{i}$ is completely symmetric.

Now assume by contradiction that the  rendezvous time $t$ is less than the claimed one, that is, $t < \frac{n}{2(c-1)}$. At time~$t$, both agents meet at some location $u$. Since $t\in I$, the above claim holds for $S_t$. Hence,  at time $t/c$, agent $A$ is at $\bar{u}$, the symmetric location of $u$. Since rendezvous happened at time $t$, this means that~agent~$A$ traveled from $\bar{u}$ to $u$ (i.e., a distance of $n/2$) during the time period $[\frac{t}{c}, t]$. Therefore $t(1-\frac{1}{c})c\geq n/2$, contradicting the assumption that $t < \frac{n}{2(c-1)}$. This establishes that any algorithm  requires $\frac{n}{2(c-1)}$ time.

We now show the simpler part of the theorem, namely, that the rendezvous time of any  algorithm in the white board model  is at least $\frac{n}{c+1}$.
Let us represent the cycle as the reals modulo $n$, that is, we view the cycle as the continuous collection of reals $[0,n]$, where $n$ coincides with 0. Assume that the starting point of agent $A$ is 0.
Consider the  time period $T=[0, \frac{n}{c+1}-\epsilon]$, for some small positive $\epsilon$. In this time period, agent $A$ moves a total length of less than $\frac{nc}{c+1}$.
Let $r$ (and $\ell $, correspondingly) be the furthest point from $0$ on the cycle that $A$ reached while going clockwise (or anti-clockwise, correspondingly), during that time period.
Note that there is a gap of length larger than $n-\frac{nc}{c+1}=\frac{n}{c+1}$ between $\ell$ and $r$. This gap corresponds to an arc not visited by agent $A$ during this time period. On the other hand, agent~$B$ walks a total distance of less than  $\frac{n}{c+1}$ during the time period $T$. Hence, the adversary can locate agent $B$ initially at some point in the gap between $r$ and $\ell$, such that during the whole time period $T$, agent $B$ remains in this gap. This establishes the $\frac{n}{c+1}$ time lower bound, and concludes the proof of the theorem. \qed
\end{proof}

\section{An Upper Bound for the Pebble Model}
In this section, we consider the pebble model. For this model, we identify the following algorithm which turns out to be extremely efficient, especially for small values of $c$. 

\paragraph{\underline{Algorithm $\pebble$.}} Each agent (1) leaves a pebble at its initial position and then starts walking in an arbitrary direction while counting the distance travelled. If (2) an agent reaches a location with a pebble for the first time
and (3) the distance it walked  is strictly less than $\tau:=\min\{n/2,n/c\}$, then (4) the agent turns around and walks continuously in the other direction.\\

Note that Algorithm $\pebble$ does not assume  sense of direction, uses only one pebble and drops it only once:  at the initial position of the agent. The algorithm assumes that agents know the values of $n$ and $c$.

Note that the assumptions of the pebble model are weaker than the white board model. Hence,
in view of Theorem~\ref{lower}, the following theorem establishes a tight bound for the case where $c\leq 2$, a $(2-\frac{2}{c})$-approximation for the case $2<c\leq 3$, and a $(c+1)/c$-approximation for the case $c>3$.\\
\begin{theorem}
 The rendezvous time of Algorithm $\pebble$ is $\max\{\frac{n}{2(c-1)}, \frac{n}{c} \}$. 
\end{theorem}
\begin{proof}
Consider Algorithm $\pebble$. First note that if both agents happen to start walking in opposite directions (due to lack of sense of direction), then they walk until they meet. In this simple case, their relative speed is $c+1$, hence rendezvous happens in time at most $\frac{n}{c+1}<\max\{\frac{n}{2(c-1)}, \frac{n}{c} \}$. For the remaining of the proof, we  consider the case that both agents start walking at the same direction, which is without loss of generality, the clockwise direction. Let  $d$ be the initial clockwise distance from $A$ to~$B$, and recall that $\tau=\min\{n/2,n/c\}$. Consider three cases.

\begin{enumerate}
\item 
$d=\tau$ (see Figure \ref{fig:upper1}).\\
Here, no agent turns around. In other words, they behave  exactly as in $\DR$. If $d=n/2$, Observation~\ref{DR} implies that the rendezvous time is $\frac{n}{2(c-1)}$. Otherwise, $c>2$ and $d=n/c$.
By Observation~\ref{DR},  the rendezvous is reached earlier, specifically, by time  $\frac{d}{c-1}=\frac{n}{c(c-1)}$.
\begin{figure}[ht]
\includegraphics[width=0.3\textwidth]{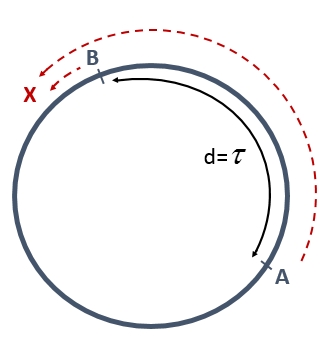}
\centering
\caption{If $d = \tau$ both agents won't turn and rendezvous will take 
$\frac{d}{c-1} < \frac{n}{2(c-1)}$ time.}
\label{fig:upper1}
\end{figure}

\item $d < \tau $ (see Figure \ref{fig:upper2}).\\
In this case, Agent $A$ will reach $B$'s starting point $v_B$, at time $d/c$, before $B$ reaches $A$'s starting point $v_A$. Moreover, agent $B$ does not turn, since its initial distance to $A$'s starting point is at least $\tau$.
At time $d/c$,
agent~$B$ is at distance~$d/c$ clockwise from $v_B$. By the algorithm, Agent $A$ then turns
and walks anti-clockwise. The anti-clockwise distance from $A$ to $B$ is then $n- d/c$.
Their relative speed is $c+1$. Hence, they will rendezvous in an additional time of
$\frac{n - d / c}{1+c}$, since no agent may turn around after time $d/c$.
Hence, the total time for reaching rendezvous is at most
$$
d / c + \frac{n - d / c}{1+c}~=~ \frac{d+n}{1+c}.
$$
\begin{figure}[ht]
\includegraphics[width=0.3\textwidth]{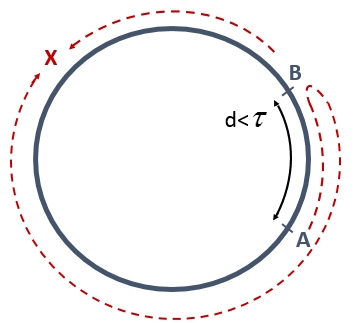}
\centering
\caption{If $d < \tau$ only $A$ will turn and rendezvous will take $\frac{n+d}{c+1}$ time.}
\label{fig:upper2}
\end{figure}

This function is maximized when $d = \tau$ where it is $\frac{\tau+n}{1+c}$.
Now, if $c\leq 2$, we have $\tau=n/2$ and the rendezvous time is therefore $\frac{3n}{2(1+c)}$.
Since $c\leq 2$, the later bound is at most $\frac{n}{2(c-1)}$.
On the other hand, if $c> 2$, we have $\tau=n/c$ and the rendezvous time  is $n/c$.
\item
$d > \tau $.\\
In this case, $A$ doesn't turn when it hits $B$'s initial position. Consider the following sub-cases.

\begin{enumerate}
\item (see Figure \ref{fig:upper3a})
The agents meet before $B$ reaches $A$'s initial position.\\
In this case, the rendezvous time (as in $\DR$) is $d/(c-1)$. On the other hand, the rendezvous time  is at most $n-d$ since $B$ did not reach $A$'s initial position. So
$d/(c-1)\leq n-d$. A simple calculation now  implies that the rendezvous time $d/(c-1)$ is at most ${n}/{c}$.

\begin{figure}[ht]
\includegraphics[width=0.3\textwidth]{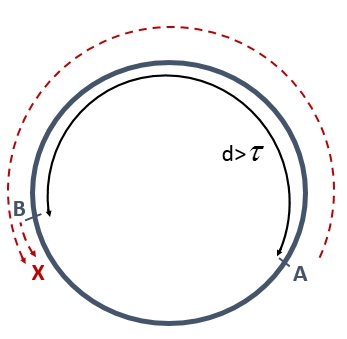}
\centering
\caption{If $d > \tau$ and rendezvous happens before $B$ reaches $A$'s pebble, it will take $\frac{d}{c-1}$ time.}
\label{fig:upper3a}
\end{figure}

\item (Figure \ref{fig:upper3b})
Agent $B$ reaches $A$'s initial position before rendezvous.\\
In this case, Agent $B$ walks for $d'=n-d$ time to first reach $A$'s initial position.
We first claim that $d'<\tau$. One case is that $c\leq 2$, and thus, $\tau=n/2$. Since $d>\tau$, we have $d'<n/2=\tau$. The other case is that
 $c>2$, so $\tau=n/c$. We claim that also in this case, we have $d'<\tau$. Otherwise, we would have had $d'\geq n/c$, which would have meant that the faster agent $A$ would have had, at least, $n/c$ time
before $B$ reached the initial position of $A$. So much time would have allowed it to cover the whole cycle. This contradicts the assumption that $B$ reached $A$'s initial position before rendezvous.
This establishes the fact that, regardless of the value of $c$, we have
$d'<\tau$.
This fact implies that
 when agent $B$ reaches $A$'s initial position, it turns around and both agents go towards each other. By the time $B$ turns around, $A$ has walked a distance of $cd'$. Hence,  at that point in time, they are $n-cd'$ apart. This implies to the following rendezvous time:

$$
d' + \frac{n - cd'}{1+c}~=~\frac{2n-d}{1+c}.
$$
Now recall that we are in the case that agent $B$ reaches $A$'s initial position before they rendezvous. This implies that $n-d<{n}/{c}$.
Hence, the running time is at most
$$
\frac{2n-d}{1+c}<\frac{n+\frac{n}{c}}{1+c}=\frac{n}{c}.
$$
\begin{figure}[ht]
\includegraphics[width=0.3\textwidth]{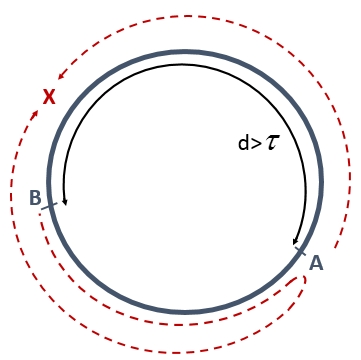}
\centering
\caption{If $d > \tau$ and $B$ reaches $A$'s pebble, then it will always turn and rendezvous will take 
$\frac{2n-d}{1+c}$ time.}
\label{fig:upper3b}
\end{figure}

\end{enumerate}

\end{enumerate}
\qed
\end{proof}

\section{Rendezvous without Communication}
In this section, we consider the case that agents cannot use marks (e.g., pebbles) to mark their location. More generally, the agents cannot communicate in any way (before rendezvous). 

In Section \ref{sec:lower-no-comm} we by show a lower bound of $\frac{cn}{c^2 - 1}$ for the rendezvous time of any algorithm, even assuming sense of direction.
Then, in Section \ref{sec:up-no-comm} we show that when assuming a sense of direction this latter bound is tight. Finally, we conclude by showing several upper bounds for the case where there is no sense of direction. 

\subsection{A Lower Bound for the Case Without Communication}\label{sec:lower-no-comm}
\begin{theorem}\label{thm:lower-nopebbles}
Consider the case that agents cannot communicate at all. The rendezvous time of any algorithm
is, at least,  $\frac{cn}{c^2 - 1}$, even assuming sense of direction.
\end{theorem}
\begin{proof}
Given an algorithm, let $\hat{t}$ denote the rendezvous time of the algorithm, that is, the maximum time (over all initial placements and all cycles of length $n$) for the agents (executing this algorithm) to reach rendezvous.
Recall, in this part of the theorem, we assume that agents have sense of direction. Without loss of generality, we assume that the direction an agent starts walking is clockwise.

 Consider, first, two identical cycles $C_A$ and $C_B$ of length $n$ each.
Let us mark a location $v\in C_A$ and a location $u\in C_B$.
 Let us examine the (imaginary) scenario in which agent $A$ (respectively, $B$) is placed on the cycle $C_A$
(respectively, $C_B$) alone, that is, the other agent is not on the cycle. Furthermore, assume that agent $A$ is placed at $v$ and agent $B$
is placed at $u$. In this imaginary scenario, agents $A$ and $B$ start executing  the protocol at the same time, separately, on each
of the corresponding cycles.
 Viewing  $u$ as homologous to $v$,  a homologous location  $h(x)\in C_B$ can be defined for each location in $x\in C_A$ in a natural way (in particular, $h(v)=u$).

  For each time $t\in [0,\hat{t}]$, let $d(t)$ denote  the clockwise distance
 between the location  $b_t\in C_B$ of the slower agent  $B$ at time $t$ and the homologous location  $h(a_t)\in C_B$ of the location $a_t\in C_A$ of the faster agent $A$ at time $t$.
 Note that $d(t)$ is a real value in $[0,n)$. Initially, we assume that all reals in $[0,n)$ are colored {\em white}.
 As time passes, we {\em color} the corresponding distances by black, that is,
at every  time $t$, we  color the distance $d(t)\in [0,n)$  by black. Note that the size of the set of black distances is monotonously non-decreasing with time.

 We first claim that,
 by time $\hat{t}$,
 the whole range $[0,n)$ is colored black. To prove by contradiction, assume that there is a real $d\in [0,n)$ that remains white.
 Now, consider the execution on a single cycle of length $n$, where agent $A$ is initially placed
 at anti-clockwise distance $d$ from agent $B$. In such a case, rendezvous is not reached by time $\hat{t}$, contrary to the assumption that it is. This implies that the time $T$ it takes until all reals in  $[0,n)$ are colored black in the imaginary scenario, is a lower bound on the rendezvous time, that is, $T\leq \hat{t}$.
It is left to analyze the time $T$.

With time, the change in the distance $d(t)$ has to follow two rules:
\begin{itemize}
\item
{\bf R1.} After one time unit, the distance can change by at most $1+c$  (the sum of the agents'  speeds).
\item
{\bf R2.} At time $x$, the distance, in absolute value is, at most, $x(c-1)$.
\end{itemize}
 To see why Rule R2 holds, recall that the programs of the agents are identical. Hence, if  agent $A$ is at some point $a\in C_A$, after completing $x$ units of distance, then agent $B$ is at point $b=h(a)$ when completing its own $x$ units of distance.
This happens for agent $B$ at time $x$ and for agent $A$ at time $x/c$. Since $A$'s speed is $c$, the maximum it can get away from point $a$ during the time period from $x/c$ until $x$ is $c(x-x/c) = x(c-1)$.

Consider the path $P:=d(t)$ covering the range  $[0,n)$ in $T$ time.
  First, note that this path $P$ may go several times through the zero point
  (i.e., when $d(t)=0$).
At a given time $s$, we say that the path is {\em on the right} side, if the last time  it left the zero point before time $s$ was while going towards 1. Similarly, the path is {\em on the left}, if the last time it left the zero point before time $s$ was while going towards $n-1$.
Let $x$ denote the largest point on the range  $[0,n)$ reached by the path while the path was on the right side. Let $y=n-x$.
By time $T$, path $P$ had to go left to distance $y$ from the zero point.
Assume, w.l.o.g. that $x<y$. (In particular, $x$ may be zero.)
The fastest way to color these two points (and all the rest of the points, since those lie between them), would be to go from zero to the right till reaching $x$, then return to zero and go to distance $y$ on the left.
Hence, $T$ will be at least:
$
T \geq \frac{x}{c-1} + \frac{n}{c+1}.
$
Indeed, Rule R2, applied to the time of reaching distance $x$, implies the first term above. The second term uses Rule R1 to capture the time
that starts with the distance reaching $x$, proceeds with the distance reaching the zero point and ends when the distance reaches $y$ when going left.
Since $c>1$, we obtain
\begin{equation}\label{eq:small}
x ~ \leq ~ \left(T - \frac{n}{c+1}\right)(c-1).
\end{equation}
On the other hand, applying Rule R2 to the final destination  $y$, we have
$
T(c-1) \geq y= n - x.
$
This implies that:
\begin{equation}\label{eq:large}
x ~ \geq ~ n - T(c-1).
\end{equation}
Combining Equations \ref{eq:small} and \ref{eq:large}, we get $T  \geq \frac{cn}{c^2 - 1}$, 
establishing the theorem.
\qed
\end{proof}

\subsection{Upper Bounds for the Case Without Communication}\label{sec:up-no-comm}
To establish the upper bounds, we consider a revised version of the $\DR$ algorithm, called $\Turn(k)$, which consists of two stages.
\paragraph{\underline{Algorithm $\Turn(k)$.}}
 At the first stage, the agent walks over its own clockwise direction for $k$ units of distance. Subsequently (if rendezvous hasn't occurred yet), the agent executes the second stage of the algorithm: it turns around and goes in the other direction until rendezvous.

\subsubsection{Assuming Sense of Direction.}
We first consider the case that agents have a sense of direction.  Recall that for this case Theorem \ref{thm:lower-nopebbles} establishes a lower bound of $\frac{cn}{c^2 - 1}$. We now show this this bound is tight.

\begin{theorem}\label{thm:upper-nopebbles-sense}
Assuming sense of direction, the rendezvous time of algorithm $\Turn(\frac{cn}{c^2 - 1})$ is $\frac{cn}{c^2 - 1}$.
\end{theorem}

\begin{proof}
Note that algorithm $\DR$ does not assume sense of direction, and its complexity is the one required by the third part of the theorem for the case $c>3$.
Recall that $\Turn(k)$ consists of two stages. At the first stage, the agent walks over its own clockwise direction for $k$ units of distance. Subsequently (if rendezvous hasn't occurred yet), the agent executes the second stage of the algorithm: it turns around and goes in the other direction until rendezvous.

Consider now Algorithm $\Turn(k)$, with parameter $k=\frac{cn}{c^2 - 1}$.
Since we assume sense of direction, both agents walk initially in the same direction (clockwise). Assume by contradiction that rendezvous hasn't occurred by time $k$.
By that time, agent $B$ travelled $k$ units of distance. Agent $A$ has travelled those $k$ units of distance
by time ${k}/{c}$. At that time, agent $A$ turns.
(However, agent $B$ will turn only at time $k$).
Hence, between those two turning times, there is a time duration $k(1 - \frac{1}{c})$
where the two agents walk towards each other.
 Hence, at each time unit they shorten the distance between them by $1+ c$.
 Let $d'$ denote the maximum distance between the agents at time ${k}/{c}$.
 It follows that $d'>k(1 - \frac{1}{c})({1+c})=n$.  A contradiction. \qed
\end{proof}

\subsubsection{Operating Without a Sense of Direction.}
\begin{theorem}\label{thm:upper-nopebbles-no-sense}
Without assuming sense of direction,  the following rendezvous time can be achieved:
\begin{enumerate}
\item
$\frac{cn}{c^2 - 1}$, for $c\leq 2$ (achieved by algorithm $\Turn(\frac{cn}{c^2 - 1})$),
\item
$\frac{2n}{c + 1}$, for $2< c\leq 3$ (achieved by algorithm $\Turn(\frac{cn}{c + 1})$),
\item
$\frac{n}{c - 1}$, for $c>3$ (achieved by algorithm $\DR$).
\end{enumerate}
\end{theorem}
\begin{proof}
Let us first consider the case $c\leq 2$. Here we apply the same algorithm  above, namely Algorithm $\Turn(k)$, with parameter $k=\frac{cn}{c^2 - 1}$.
As proved before, if the two agents happen to start at the same direction then rendezvous occurs by time $\frac{cn}{c^2 - 1}$. Hence, let us consider the case that both agents
walk initially at opposite directions. Assume by contradiction, that  rendezvous hasn't occurred by time $k/c$.
In this case, by time $k/c$ the faster agent $A$ walked $k$ units of distance toward $B$, and the slower agent $B$ walked $k/c$ units of distance towards $A$. Hence, the initial distance between them must be greater than $k(1+1/c)=n/(c-1)>n$, a contradiction. This proves the first item of the Theorem.

Note that Algorithm $\DR$ establishes the third item of the Theorem. Hence, it is left to prove the second item, namely, the case $2<c\leq 3$.
For this case, we apply Algorithm $\Turn(k)$, with parameter $k=\frac{cn}{c + 1}$. First note, if the two agents happen to start walking towards each other they continue to do that until time
$k/c=\frac{n}{c + 1}$, hence they would meet by this time. Therefor, we may assume that initially, both agents start in the same direction. In this case, if rendezvous hasn't occurred by time $k/c$, then at this time agent $A$ turns around, and both agents walk towards each other for at least $k(1-1/c)$ more time (the time when agent $B$ is supposed to turn). Since $c>2$, we have $k(1-1/c)>k/c$, and hence, from the time agent $A$ turns, both agents walk towards each other for at least $\frac{k}{c}=\frac{n}{c + 1}$ time, which means they must meet by this time. Altogether, the time to rendezvous is $\frac{2k}{c}=\frac{2n}{c + 1}$, as desired.
\qed \end{proof}

\section{Discussion and Future Work}
We show how some form of  asynchrony could be useful for solving a symmetry breaking problem efficiently.
 Our study could be considered as a first attempt to harness the (unknown) heterogeneity between individuals in a cooperative population towards more efficient functionality.

There are many natural ways of further exploring this idea in future work.
First, we have studied the exploitation of asynchrony for a specific kind of problems. It seems that it can be useful for other symmetry breaking problems as well.
Another generalization:
the ``level'' of asynchrony considered in this paper is very limited:  the ratio $c$ between the speeds of the agents is the same throughout the execution, and is known to the agents. Exploiting a higher level of asynchrony should also be studied, for example,
the case that the speed difference is stochastic and changes through time.

Our main interest was the exploitation of asymmetry, rather than the specific problem of rendezvous. Still,
 even for the rendezvous problem, this current study leaves many open questions. First, even for the limited case we study, not all our bounds are completely tight, and it would be interesting to close the remaining gaps between our lower and upper bounds (these gaps hold for some cases when $c>2$). In addition, it would be interesting to study further the uniform case \cite{KSV11}, in which $n$ and $c$ are not known to agents. Another direction is to generalize the study to multiple agents (more than two, see, e.g., \cite{DFPS03,multiple-r}) and to other topological structures. This would also allow one to study interactions between various means of breaking symmetry (such as different speeds together with different locations on a graph).

\section{Acknowledgement}
O.F., incumbent of the Shlomo and Michla Tomarin Career Development Chair, was supported by the Clore Foundation, the Israel Science
Foundation (FIRST grant no. 1694/10) and the Minerva Foundation.
A.K. was supported  by the ANR project DISPLEXITY, and by the INRIA project GANG. 
S.K. was supported in part by the ISF and by the Technion TASP center.

This work has received funding from the European Research Council (ERC) under the European Union's Horizon 2020 research and innovation program (grant agreement No 648032).

\end{document}